\documentclass[traditabstract]{aa}
\usepackage{graphicx}
\usepackage{txfonts}
\usepackage{natbib}

\newcommand{\Msun}{\mbox{$M_{\odot}$}}

\begin{document}
\title{Accretion in the detached post-common-envelope binary LTT 560
\thanks{Based on observations made at ESO telescopes (079.D-0276)}}

\author{
C.~Tappert\inst{1},
B.~T.~G\"ansicke\inst{2},
L.~Schmidtobreick\inst{3},
T.~Ribeiro\inst{4}}

\authorrunning{C. Tappert et al.}
\titlerunning{Accretion in the detached PCEB LTT 560}

\offprints{C. Tappert}

\institute{
Departamento de F\'{\i}sica y Astrof\'{\i}sica, Universidad de Valpara\'{\i}so,
Av. Gran Breta\~na 1111, Valpara\'{\i}so, Chile\\
\email{ctappert@dfa.uv.cl}
\and
Department of Physics, University of Warwick, Coventry CV4 7AL, UK\\
\email{Boris.Gaensicke@warwick.ac.uk}
\and
European Southern Observatory, Casilla 19001, Santiago 19, Chile\\
\email{lschmidt@eso.org}
\and
Departamento de F\'{\i}sica, Universidade Federal de Santa Catarina, Campus 
Trindade, 88040-900 Florian\'opolis, SC, Brazil\\
\email{tiago@astro.ufsc.br}}

\date{Received xxx; accepted xxx}

\abstract{
In a previous study, we found that the detached post-common-envelope binary
LTT 560 displays an H$\alpha$ emission line consisting of two anti-phased
components. While one of them was clearly caused by stellar activity from
the secondary late-type main-sequence star, our analysis indicated that the 
white dwarf primary star is potentially the origin of the second component. 
However, the low resolution of the data means that our interpretation remains
ambiguous. We here use time-series UVES data to compare the radial velocities 
of the H$\alpha$ emission components to those of metal absorption lines from 
the primary and secondary stars. We find that the weaker component most 
certainly originates in the white dwarf and is probably caused by accretion. 
An abundance analysis of the white dwarf spectrum yields accretion rates that 
are consistent with mass loss from the secondary due to a stellar wind. The 
second and stronger H$\alpha$ component is attributed to stellar activity 
on the secondary star. An active secondary is likely to be present because
of the occurrence of a flare in our time-resolved spectroscopy. Furthermore, 
Roche tomography indicates that a significant area of the secondary star on 
its leading side and close to the first Lagrange point is covered by star 
spots. Finally, we derive the parameters for the system and place it in
an evolutionary context. We find that the white dwarf is a very slow rotator,
suggesting that it has had an angular-momentum evolution similar to that of
field white dwarfs. We predict that LTT 560 will begin mass transfer via 
Roche-lobe overflow in $\sim$3.5 Gyrs, and conclude that the system is 
representative of the progenitors of the current population of cataclysmic 
variables. It will most likely evolve to become an SU UMa type dwarf nova.
}


\maketitle

\section{Introduction}

\defcitealias{tappertetal07-1}{Paper I}%
Cataclysmic variables (CVs) are thought to form from detached
white dwarf (WD) / main-sequence star (MS) binaries that have experienced a 
common-envelope (CE) phase \citep[e.g., ][]{taam+ricker10-1,webbink08-1}. In 
these ``post-common-envelope binaries" (PCEBs), the white dwarf represents the 
more massive component in the system and is therefore called the primary, 
while the usually late-type (K--M) main-sequence star is known as the 
secondary. After the end of the CE phase, angular-momentum loss due to 
magnetic braking and/or gravitational radiation continues to decrease the 
separation between the two stars. This eventually brings the secondary's Roche 
lobe into contact with the stellar surface, thus initiating stable 
mass-transfer via Roche-lobe overflow and the semi-detached CV phase 
\citep[][ and references therein]{ritter08-1}.

There is growing evidence that the accretion of material from the secondary 
star does not start with Roche-lobe overflow. The discovery of so-called
``low accretion rate polars'' \citep{schwopeetal02-1}, which are likely 
progenitors of magnetic CVs \citep{schmidtetal05-1,schmidtetal07-1}, and the 
detection of metal absorption lines in the UV spectra of non-magnetic PCEBs
\citep[e.g., ][]{kawkaetal08-1}, indicate that wind accretion from the active 
chromosphere of the secondary star is a common phenomenon in detached, 
short-period, WD/MS binaries.

\citet[][ hereafter Paper I]{tappertetal07-1} found the H$\alpha$ emission
line in the detached PCEB \object{LTT 560} to be a combination of 
two components. The radial velocity variations in the stronger H$\alpha$ 
component agree well with those shown by the TiO absorption, and was thus 
identified as originating in the secondary star. The radial velocities of the 
weaker component showed an -- within the errors -- anti-phased behaviour with 
respect to the secondary star exhibiting a lower amplitude, and therefore had 
to be produced on the side of the centre-of-mass opposite to the secondary. 
However, the low spectral resolution of the data, and the contamination of 
the white-dwarf Balmer absorption lines with emission cores from the 
secondary, impeded velocity measurements of other, unambiguously intrinsic, 
spectral features from the white dwarf. Thus, the true origin of the weak 
emission component remained unresolved. The low temperature of the white dwarf 
in LTT 560 ($T \sim 7500~\mathrm{K}$, Paper I) and the evidence of ongoing 
accretion imply that there is a high probability that there are narrow metal 
absorption lines in the white-dwarf spectrum 
\citep{zuckermanetal03-1}. The radial velocity curve of such lines should be 
undisturbed and thus faithfully track the motion of the white dwarf, 
motivating the high-resolution study presented in this paper.

\section{Observations and data reduction}

LTT 560 was observed on August 16, 2007 using UVES on UT1 (ANTU) at ESO
Paranal. A total of 38 Echelle spectra was taken over 7.3 h in one blue 
(3260--4528 {\AA}) and two red (5681-7519 {\AA}, 7661-9439 {\AA}, hereafter
"lower" and "upper" red spectrum, respectively) spectral 
ranges, with resolving power $\sim$70\,000. The exposure time per spectrum 
was 600 s, which corresponds to roughly 0.05 orbital cycles. The sequence of 
time-resolved spectra had to be interrupted for about 45 min when the 
object was close to the zenith, where the VLT cannot observe.

The data reduction was performed using Gasgano and the UVES pipeline (version
3.9.0) in a step-by-step mode. This included bias and flat correction, 
wavelength calibration with a ThAr lamp, and flux calibration using the
standard star LTT 7987. The response function proved rather unsatisfactory, 
with the flux-calibrated spectra still containing several "order wriggles". 
However, since this bears no relevance to the work described in this paper, 
we did not attempt to resolve this issue.

\section{Results}

\begin{figure}
\includegraphics[width=\columnwidth]{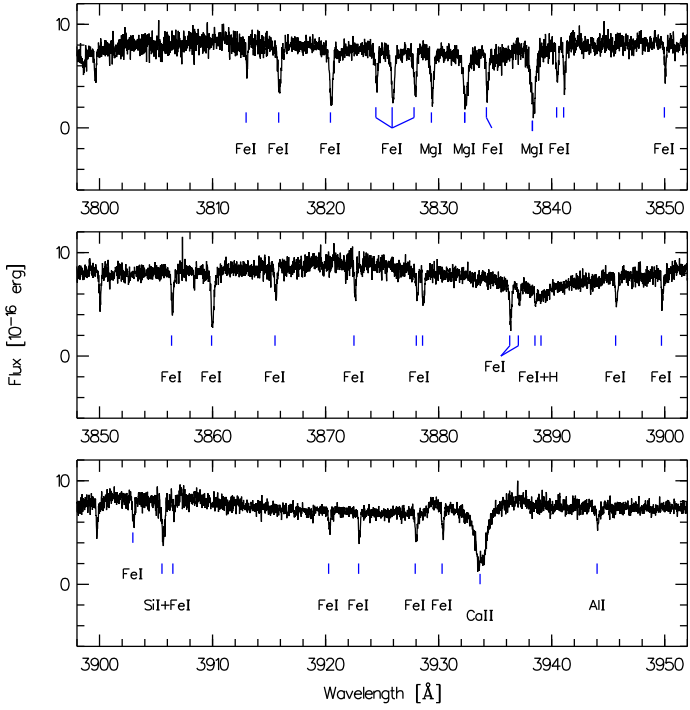}
\caption[]{Selected ranges of the blue spectrum with line identifications.
This average spectrum has been produced by combining 38 individual spectra 
that had each been corrected for the radial velocity variations of the white 
dwarf.}
\label{bluespec_fig}
\end{figure}

\begin{figure}
\includegraphics[width=\columnwidth]{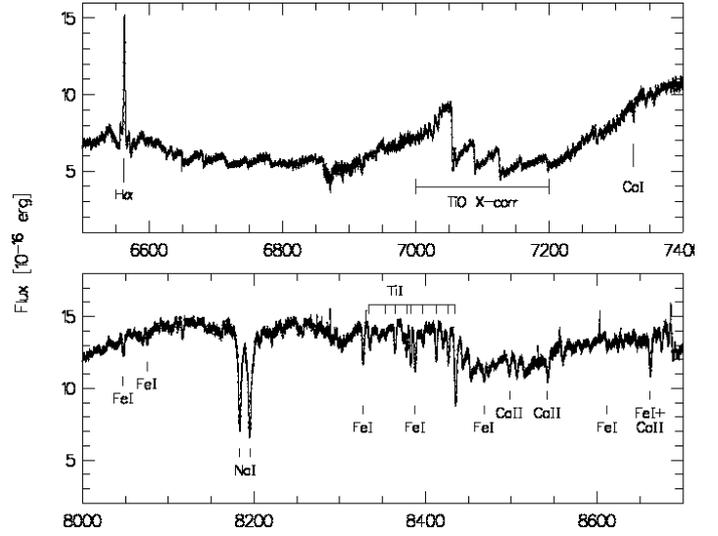}
\caption[]{Selected ranges of the average red spectra. The respective
individual spectra have been corrected for the radial velocities of the
secondary star. The spectra are not corrected for telluric
contamination. Indicated are the H$\alpha$ emission line, the 
spectral range that has been used for the cross-correlation, and the 
absorption lines that could be unambiguously identified. 
}
\label{redspec_fig}
\end{figure}

In Figs.\,\ref{bluespec_fig} and \ref{redspec_fig}, we present selected
wavelength ranges of the averaged blue and red spectra. For the computation of
both spectra, the individual data were corrected for the motion of
the respective components based on the radial velocity parameters derived
in Section \ref{rv_sec}. Thus, Fig.\,\ref{bluespec_fig} shows the blue
spectrum in the rest frame of the white dwarf, while Fig.\,\ref{redspec_fig}
gives the red spectra in the rest frame of the late M dwarf. Absorption
features of both stellar components were identified using the ILLSS
catalogue \citep{coluzzi99-1}.

The two red spectra are dominated by the features of the secondary star, which
are mainly molecular TiO bands, but also several atomic absorption lines such 
as Ca\,{\sc I} $\lambda$7326 (in the "lower" red spectrum), the 
Na\,{\sc I} $\lambda\lambda$8183/8194 doublet, K\,{I}, Fe\,{\sc I}, Ti\,{I},
and Ca\,{II} (in the "upper" red spectrum). As shown in 
\citetalias{tappertetal07-1}, the red
spectrum is consistent with an M5--6V secondary star.
The blue portion of the spectra presents a plethora of narrow metal
absorption lines: these are mostly of Fe\,{\sc I}, but also Ni\,{\sc I}, 
Si\,{\sc I}, Mg\,{\sc I}, Al\,{\sc I}, Ca\,{\sc I}, and Cr\,{\sc I} can be 
identified. In addition, there are the broader lines of Ca\,{\sc II}, and 
H\,{\sc I}. In Section \ref{wdsp_sec}, we use the absorption lines of the 
white dwarf to estimate the accretion rate.

\subsection{Radial velocities\label{rv_sec}}

\begin{table}
\caption[]{Radial velocity parameters.}
\label{rvpars_tab}
\begin{tabular}{llll}
\hline\hline
Feature & $K$ & $\gamma$ & $\varphi$ \\
        & [km s$^{-1}$] & [km s$^{-1}$] & [orbits] \\
\hline
H$\alpha_\mathrm{MS}$\tablefootmark{a} & 231.50$\pm$0.51 & 35.01$\pm$0.46 
& $-$0.00271(65)\\
Na\,{\sc I} $\lambda$8183 & 233.72$\pm$0.33 & 35.59$\pm$0.24 & 0.00011(24) \\
Na\,{\sc I} $\lambda$8195 & 228.92$\pm$0.29 & 37.96$\pm$0.22 & $-$0.00019(22)\\
TiO cc+$\lambda_0$\tablefootmark{b}    & 231.17$\pm$0.19 & 37.18$\pm$1.28 
& 0.00009(14) \\
H$\alpha_\mathrm{WD}$\tablefootmark{c} &  72.70$\pm$0.15 & 55.69$\pm$0.14 
& 0.50198(62) \\
metal$_\mathrm{av}$\tablefootmark{d}   &  72.63$\pm$0.88 & 54.24$\pm$0.88 
& 0.5009(20)  \\
metal cc+$\lambda_0$\tablefootmark{e}  &  72.62$\pm$0.16 & 55.20$\pm$0.74 
& 0.50018(37) \\
\hline
\end{tabular}
\tablefoot{
\tablefoottext{a}{broad H$\alpha$ component excluding conjunction spectra.}
\tablefoottext{b}{$K$ and $\varphi$ via cross-correlation of the range 
7000--7200\,{\AA}, $\gamma$ from the position of 15 red absorption lines.}
\tablefoottext{c}{narrow H$\alpha$ component excluding conjunction spectra.}
\tablefoottext{d}{average of the radial velocity parameters of 16 blue metal 
lines.}
\tablefoottext{e}{$K$ and $\varphi$ via cross-correlation of the range 
3810--3870\,{\AA}, $\gamma$ averaged from the positions of 31 metal lines.}
}
\end{table}

\begin{figure}
\includegraphics[angle=-90,width=\columnwidth]{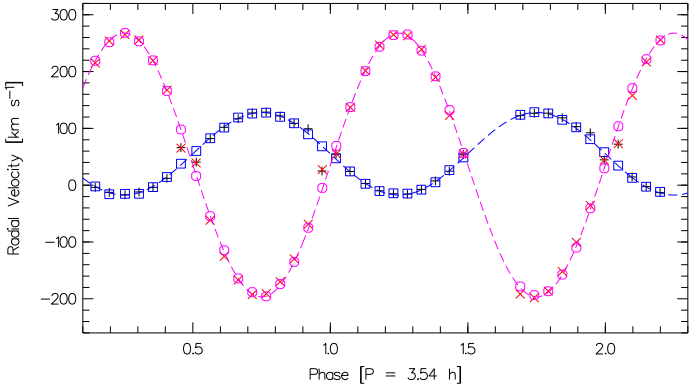}
\caption[]{Radial velocities of several spectral features plotted against
orbital phase. The latter are presented in sequence, i.e.\,corresponding to
their time of measurement. The plot shows the radial velocities for the
broad and narrow H$\alpha$ components (marked by slanted and straight crosses, 
respectively), and the TiO and metal cross-correlation velocities
(circles and squares, resp.). The plot also includes the radial velocity 
curves calculated from the adapted average radial velocity parameters.
}
\label{rvs_fig}
\end{figure}

\begin{figure}
\centering
\includegraphics[width=0.7\columnwidth]{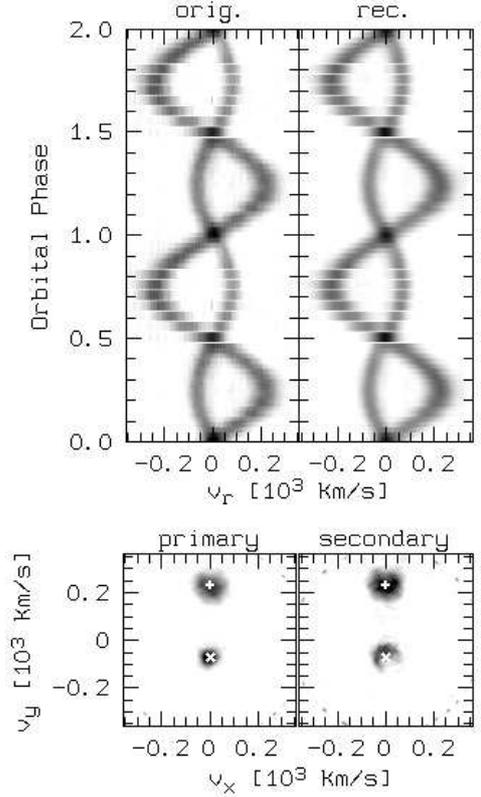}
\caption[]{Trailed spectra and Doppler maps. The {\em lower left plot} 
presents the Doppler map corrected for $\gamma_\mathrm{WD}$, the {\em lower 
right plot} the corresponding one corrected for $\gamma_\mathrm{MS}$. The 
slanted and the straight cross mark the calculated positions of the stars, 
i.e.\, $K_\mathrm{WD}$ and $K_\mathrm{MS}$, respectively. The upper plots show 
the trailed spectra for $\gamma_\mathrm{WD}$. The {\em left spectrum} 
represents the original data, the {\em right spectrum} is the reconstructed 
data from the Doppler map. The data are repeated for a second orbital cycle 
for clarity.
}
\label{doppler_fig}
\end{figure}

\begin{table}
\caption[]{System parameters of LTT 560.}
\label{syspars_tab}
\begin{tabular}{@{\extracolsep{0pt}}lllllll}
\hline\hline
WD & $P_\mathrm{orb}$ & $M_\mathrm{WD}$ & $M_\mathrm{MS}$ & $q$ & $i$ 
& $a$ \\
   & [h]              & [$M_\odot$]     & [$M_\odot$]     &     & [$^\circ$]
& [$R_\odot$] \\
\hline
He & 3.54(7) & 0.46(1) & 0.146(4) & 0.314(1) & 63(2) & 1.00(1) \\
CO & 3.54(7) & 0.44(1) & 0.139(4) & 0.314(1) & 65(2) & 0.98(1) \\
\hline
\end{tabular}
\tablefoot{%
Orbital period, masses, mass ratio, inclination, and binary separation for a 
He- and a CO-core white dwarf primary assuming a $10^{-5}\Msun$ hydrogen 
envelope.
}
\end{table}

We recall that the main motivation of the present work was to determine the 
origin of the two H$\alpha$ emission components. This can be achieved by 
calculating the parameters of their respective radial velocity variations
\begin{equation}
\label{rv_eq}
v_r(t) = \gamma + K \sin [2 \pi (t-T_0)/P_\mathrm{orb}],
\end{equation}
and by comparing them to the corresponding parameters of intrinsic
stellar absorption features. We therefore measured the radial velocities of
the following lines and spectral regions:
\smallskip\noindent\\
(1)~\,The two H$\alpha$ emission components were fitted individually with
a single Gaussian. For the calculation of the parameters, we excluded
the data close to superior and inferior conjunction, when both components 
merge. 
\smallskip\noindent\\
(2)~\,By fitting single Gaussians again, the velocities for 29 metal absorption
lines in the blue spectral range (in the majority Fe\,{\sc I}) were measured, 
and the fitting of sine functions according to Eq.\,\ref{rv_eq} yielded the 
corresponding parameters. The sample was subsequently restricted to 16 curves 
with a standard deviation $< 0.5~\mathrm{km/s}$ in $\gamma$ to provide average 
parameters.
\smallskip\noindent\\
(3)~\,The Na\,{\sc I} lines were also measured with single Gaussian
functions. An attempt to fit both lines simultaneously with two Gaussians
at a fixed separation was unsuccessful because of the irregular and variable
continuum.
\smallskip\noindent\\
(4)~\,Cross-correlation was performed for the spectral ranges 
3810--3870\,{\AA} (containing several narrow absorption lines)
and 7000--7200\,{\AA} (including three strong TiO absorption bands).
Lacking a proper radial velocity standard, the spectra were correlated to
one spectrum of the data set of high signal-to-noise ratio (S/N). The 
resulting velocities were then fitted according to Eq.\,\ref{rv_eq} with 
$\gamma$ set to 0. Subsequently, the individual spectra were corrected for the 
corresponding variation and averaged to yield high S/N spectra. Finally, 
$\gamma$ was determined by calculating the shift of spectral features with 
respect to their rest-frame wavelength in these average spectra. For the blue 
range, this shift was calculated as the average of the positions of 31 lines. 
For the red range, the Ca\,{\sc I} $\lambda$7326 line was used from the lower 
red spectrum and 12 lines from the upper red range.

Table \ref{rvpars_tab} summarises the resulting parameters. The zero point
of the orbital variation $T_0$ was set to the inferior conjunction of the
TiO cross-correlation velocities, yielding an ephemeris
\begin{equation}
T_0 ({\rm HJD}) = 2\,454\,329.6840(29) + 0.1475(29)~E,
\end{equation}
where $E$ is the cycle number. The value for the orbital period was
taken from \citetalias{tappertetal07-1}. 

As can be seen in Table \ref{rvpars_tab}, and is visualised in 
Fig.\,\ref{rvs_fig}, there is a close agreement between all parameters
for the respective features of both the white dwarf and the secondary star. 
We may thus calculate the weighted averages for these parameters:
\begin{eqnarray*}
\gamma_\mathrm{WD} = 55.45(34) ~\mathrm{km/s}, &
K_\mathrm{WD} = 72.66(11)~\mathrm{km/s},\\
\gamma_\mathrm{MS} = 36.51(72)~\mathrm{km/s}, &
K_\mathrm{MS} = 231.22(95)~\mathrm{km/s}.
\end{eqnarray*}
With this, we derive the mass ratio 
\begin{displaymath}
q = K_\mathrm{WD}/K_\mathrm{MS} = 0.3143(13)
\end{displaymath}
and the gravitational redshift of the white dwarf
\begin{displaymath}
v_\mathrm{gr} = \gamma_\mathrm{WD} - \gamma_\mathrm{MS} = 18.94(80)~
\mathrm{km/s}.
\end{displaymath}
These values are consistent with those determined in 
\citetalias{tappertetal07-1}
($q = 0.36(03)$, $v_\mathrm{gr} = 25(09)~\mathrm{km/s}$), but represent
a vast improvement in accuracy. 
Assuming a He-core and a CO-core white dwarf, and adopting the
mass-radius relations from \citet{paneietal00-2} for a $10^{-5}\Msun$
hydrogen envelope, the gravitational redshift corresponds to
$M_\mathrm{wd} = 0.46\pm0.12\Msun$ and $M_\mathrm{wd} =
0.44\pm0.12\Msun$, respectively.
The resulting updated system parameters (see \citetalias{tappertetal07-1} for 
details) are summarised in Table \ref{syspars_tab}.

In Fig.\,\ref{doppler_fig}, we
show the trailed spectrum and the Doppler map that was computed using
the fast maximum entropy algorithm of \citet{spruit98-1}. The flare data 
(Section \ref{flare_sec}) were excluded, since
Doppler tomography does not take into account non-orbital variations. The
resulting map represents a two-dimensional visualisation of the emission
distribution in velocity space \citep{marsh+horne88-1}. We note that the 
significant difference in the $\gamma$ velocities caused by the gravitational 
redshift of the white dwarf requires the calculation of two maps with
respective corrections. Consequently, the emission from the secondary star
appears distorted in the $\gamma_\mathrm{WD}$ corrected map, and vice versa.
We note that both the emission component from the white dwarf and the one
from the secondary star in their respective "$\gamma$ rest frames" are 
symmetrically centred on the calculated positions of the stellar components,
and that these two are the only H$\alpha$ emitters in the system, i.e.~no
accretion stream or disc is visible in H$\alpha$.

\subsection{The photospheric white dwarf spectrum\label{wdsp_sec}}

\begin{table}
\caption{\label{t-abundances} Photospheric white dwarf abundances.}
\begin{tabular}{lccc}
\hline\hline
Element & $\log[\mathrm{X/H}]_\mathrm{LTT560}$\tablefootmark{1} 
& $\log[\mathrm{X/H}]_\odot$\tablefootmark{2} 
& $\times$solar\tablefootmark{3}\\
\hline
Mg &  -6.2  & -4.4 & 0.015\\
Al &  -7.6  & -5.6 & 0.010\\
Si &  -6.2  & -4.5 & 0.018\\
Ca &  -7.5  & -5.7 & 0.015\\
Sc &  -10.9 & -8.8 & 0.008\\
Mn &  -8.2  & -6.6 & 0.018\\
Fe &  -6.3  & -4.5 & 0.018\\
Co &  -8.8  & -7.0 & 0.018\\
Ni &  -7.9  & -5.8 & 0.007\\
\hline
\end{tabular}
\tablefoot{%
Abundances determined from fitting TLUSTY/SYNSPEC models to the metal lines
detected in the average UVES spectrum of LTT\,560. \\
\tablefoottext{1}{Metal abundances relative to hydrogen, by number, for 
LTT\,560.}
\tablefoottext{2}{As (1), but for the Sun.}
\tablefoottext{3}{The abundances in LTT\,560 relative to those in the Sun.}
}
\end{table}

\begin{figure}
\includegraphics[width=\columnwidth]{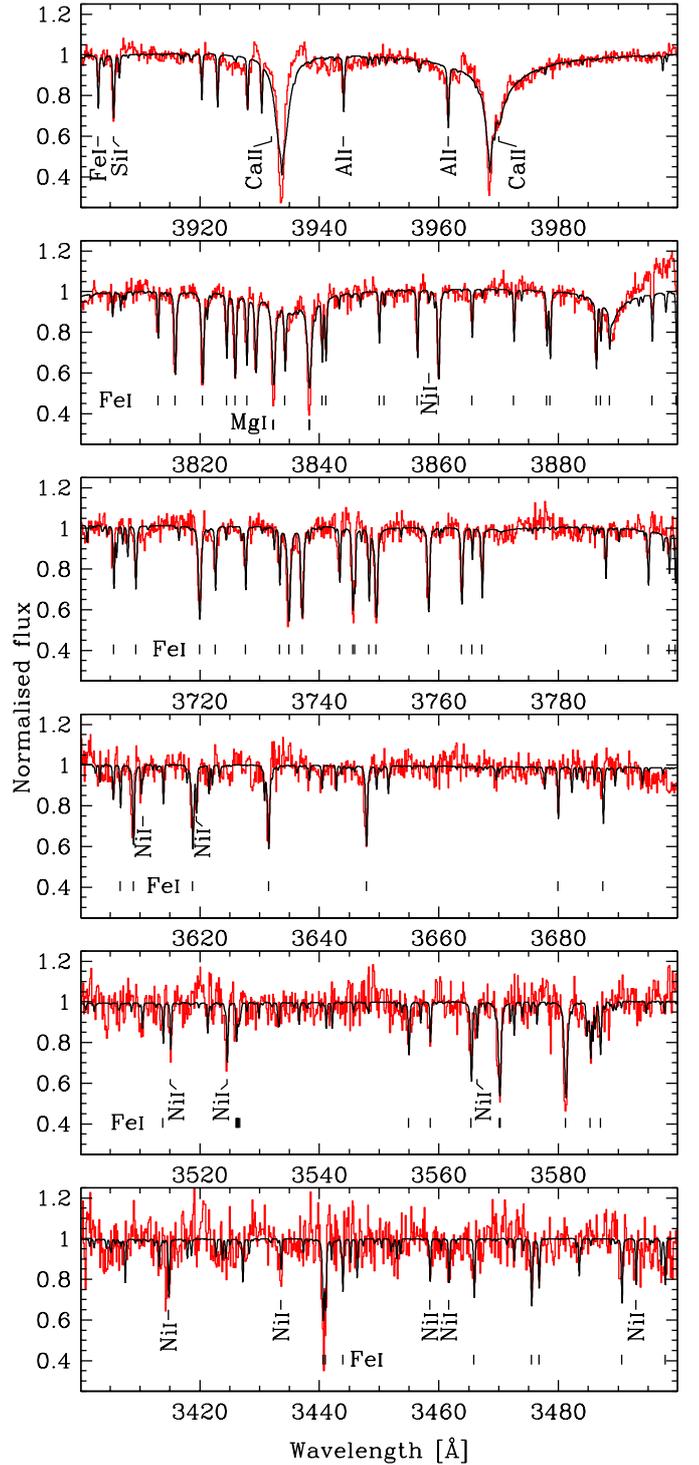}
\caption[]{\label{f-abundances} The plethora of metal absorption lines in the 
average spectrum of the white dwarf in LTT\,560 were fitted with 
TLUSTY/SYNSPEC models to determine the metal abundances in the white dwarf 
photosphere. The overplotted black line shows a model for 
$T_\mathrm{WD}=7500$\,K, $\log g=7.75$, and adopting the abundances listed in 
Table\,\ref{t-abundances}.}
\end{figure}

We used TLUSTY200 and SYNSPEC48 \citep{hubeny+lanz95-1,lanz+hubeny95-1}
along with the Kuruz line list (CD-ROM23) to model the plethora of
photospheric metal lines detected in the average UVES spectrum of the
white dwarf in LTT\,560. 
The atmospheric models were computed adopting the mixing length version ML2
and a mixing length of 0.6.
We fixed the effective temperature to
$T_\mathrm{eff}=7500$\,K, as determined from our analysis of the
near-ultraviolet to infrared spectral energy distribution of LTT\,560
\citepalias{tappertetal07-1}, and $\log g=7.75$, corresponding to the mass 
determined here. We restrict this analysis to wavelengths below 4000\,\AA, 
where the contribution of the companion star is practically zero, hence
the photospheric absorption lines from the white
dwarf are not contaminated. We identified transitions of Mg, Al, Si,
Ca, Sc, Mn, Fe, Co, and Ni, and varied the individual abundances of
these elements to achieve the closest fit to the observed line profiles
(Fig.\,\ref{f-abundances}).  The abundances determined from this fit
are reported in Table\,\ref{t-abundances}, with typical uncertainties
of 0.2\,dex.  Within these uncertainties, the observed abundance
pattern is broadly consistent with a solar element mixture at $\simeq0.015$
times the solar metal abundances \citep{asplundetal09-1}.

The relatively large abundances of metals in the photosphere of the
white dwarf clearly indicate that it is accreting from its companion
star. Doppler tomography rules out Roche-lobe overflow, hence
the origin of the accreted material is very likely to be the wind of the
companion star. Assuming that the white dwarf photosphere is in a
steady-state of accretion-diffusion, we can use the determined metal
abundances to infer the accretion rate from
\begin{equation}
\dot M = \frac{qM_\mathrm{WD} X}{\tau_\mathrm{D} X_\mathrm{acc}},
\end{equation}
where $q$ is the mass fraction of the convection zone (in which the
accreted material is mixed), $\tau_\mathrm{D}$ is the diffusion timescale on 
which accreted metals drop out of the convection zone, and
$[X/H]$ and $[X/H]_\mathrm{acc}$ are the abundances observed in the
white dwarf atmosphere and in the accreted material, respectively
\citep{dupuisetal93-1, koester+wilken06-1}. We assume that the white
dwarf in LTT\,560 accretes solar abundance material, and estimate $q=-8.2$
from the plots in
\citet{althaus+benvenuto98-1}. \citet{koester+wilken06-1} list the
diffusion timescales for Ca, Mg, and Fe for a wide range of effective
temperatures and surface gravities. We interpolate their Table\,2
for $T_\mathrm{WD}=7500$\,K and $\log g=7.75$ to find
$\tau_\mathrm{D}(\mathrm{Ca})=9400$\,yr,
$\tau_\mathrm{D}(\mathrm{Mg})=9950$\,yr, and
$\tau_\mathrm{D}(\mathrm{Fe})=7460$\,yr. Using the abundances of Ca,
Mg, and Fe determined from our fits to the photospheric metal lines
(Table\,\ref{t-abundances}), we obtain three independent estimates of
the accretion rate, $\dot M=4.4\times10^{-15}\Msun\mathrm{yr^{-1}}$,
$4.5\times10^{-15}\Msun\mathrm{yr^{-1}}$, and $\dot
M=6.0\times10^{-15}\Msun\mathrm{yr^{-1}}$. Within the uncertainties of our
analysis, we conclude that the accretion rate onto the white dwarf is
$\sim 5\times10^{-15} \Msun\mathrm{yr^{-1}}$.

\subsection{Activity}

\subsubsection{\label{flare_sec}Flaring}

\begin{figure}
\includegraphics[angle=-90,width=\columnwidth]{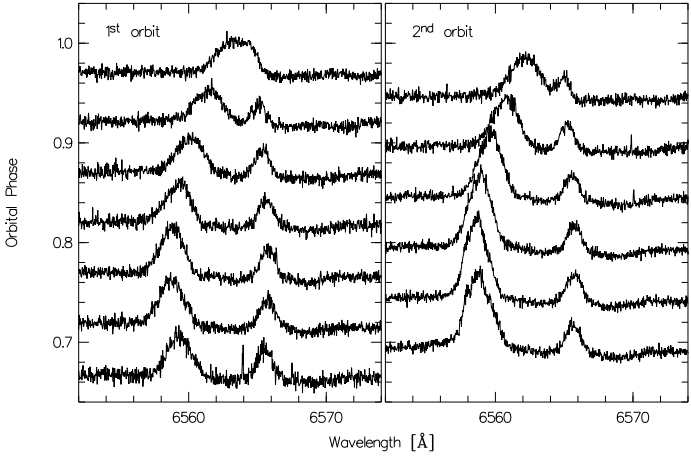}
\caption[]{Comparison of the two half orbits corresponding to the phase
intervals 0.64--1.0 (left panel) and 1.64--2.0 (right) in Fig.\,\ref{rvs_fig}. 
The H$\alpha$ component from the secondary star is clearly stronger at the
start of the second panel, while the one from the white dwarf remains unchanged.
}
\label{flareline_fig}
\end{figure}

\begin{figure}
\includegraphics[angle=-90,width=\columnwidth]{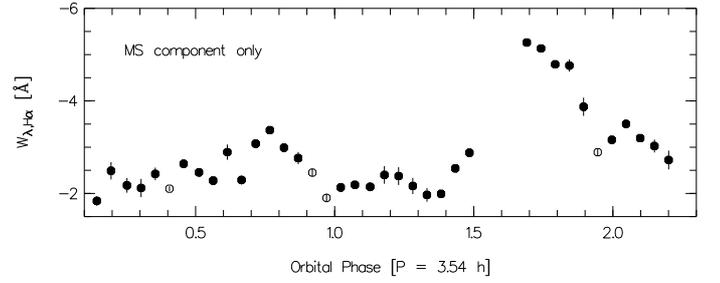}
\caption[]{Equivalent widths of the H$\alpha$ after the emission 
component of the white dwarf had been subtracted from the spectrum. Open 
circles mark spectra where the subtraction left significant negative residuals. 
}
\label{eqwha_fig}
\end{figure}

In the second part of the second orbit covered by our data, the H$\alpha$
component from the late-type star is significantly enhanced from phases 
1.7--1.9, i.e.\, immediately after the "zenith break" 
(Fig.\,\ref{flareline_fig}). 
In contrast, the white-dwarf H$\alpha$ component remains constant. For a 
quantitative analysis, we measured the equivalent width of the secondary's 
H$\alpha$ component as follows. First, we fitted single Gaussian 
functions to the line profile of the white-dwarf H$\alpha$ component in all 
spectra where this component was sufficiently isolated. Subsequently, all fits 
were subtracted from all spectra, and the quality of the subtraction was 
evaluated visually. Eight fits were found to leave sufficiently 
negligible residuals. Finally, the equivalent widths of the secondary's
H$\alpha$ component was measured in all eight sets of subtracted spectra.

For four spectra, none of the subtracted line profiles were of satisfactory
quality, containing significant negative residuals. Interestingly, three of 
those represent all the spectra in the phase interval 0.9--1.0, and one might 
speculate that at these phases (close to superior conjunction of the primary) 
the white-dwarf H$\alpha$ component is affected by some sort of obscuration. 
Nevertheless, our coverage here is insufficient for an in-depth analysis, and 
more data will be needed to confirm this apparent diminished strength of the
white-dwarf H$\alpha$ component.  

The average equivalent widths are plotted in 
Fig.\,\ref{eqwha_fig}. The secondary's H$\alpha$ component at second 
quadrature during the second observed orbit is clearly about 1.6 times 
stronger than during the first orbit. The observed maximum of this flare 
occurs at phase 1.7, but since the phase interval 1.5--1.7 was not covered
because of the zenith break it is possible that the real maximum was reached
before we recommenced observations. From phase 1.7 to the end of observations
at phase 2.2, the strength of the H$\alpha$ line had declined, but had not yet 
reached its "quiescence" value at this point. We furthermore note
that during the first observed orbit the equivalent width has a local
maximum at second quadrature, where it is about 1.4 times larger than at both
observed "unflared" first quadratures (phases 0.25 and 1.25). This indicates
that there is an active region on the leading side of the secondary star.

\subsubsection{Roche tomography}

\begin{figure}
\centering
\includegraphics[width=0.75\columnwidth]{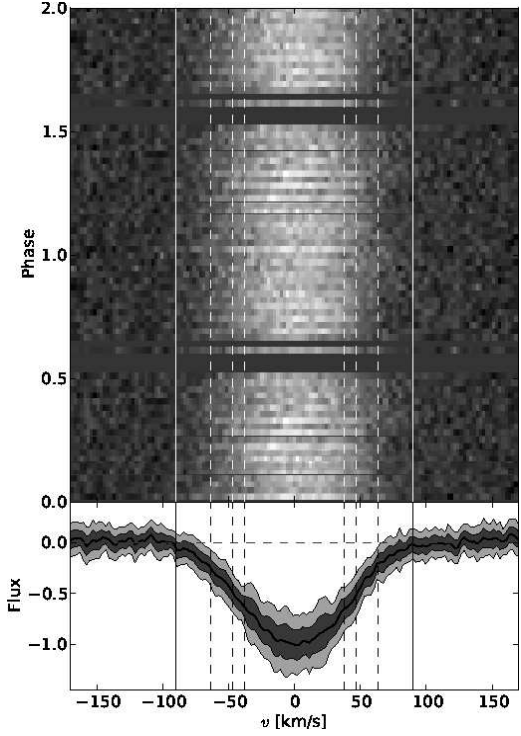}
\caption[]{Radial velocity-corrected trailed spectrum (top)
and the combined absorption line (bottom) resulting from the least square
deconvolution for the data before the zenith break. 
Dark and light grey regions on the average line profile at the bottom panel 
show the 1-$\sigma$ and 2-$\sigma$ of the standard deviation throughout the 
orbit, respectively. Solid vertical lines depict the maximum allowed radius 
for the secondary star (i.e., its Roche-lobe radius $R_{\rm L1}$). The 
vertical dashed lines mark the estimated radius $R_\mathrm{MS}$ (central lines)
and the corresponding 1-$\sigma$ confidence levels obtained from the RT by the 
$\chi^2$ landscape technique.
}
\label{lsd_fig}
\end{figure}

\begin{figure}
\centering
\includegraphics[width=0.8\columnwidth]{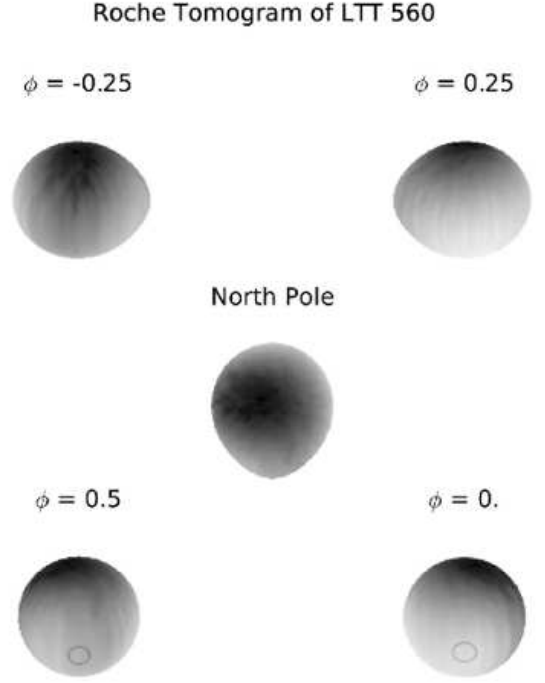}
\caption{The surface brightness distribution of the secondary star of 
LTT 560 as derived from the Roche tomography. The L1 face of the star 
is seen at phase 0.5. The circles mark a region around L1 ({\em lower left
plot:} seen from the front side; {\em lower right plot:} 
seen through the star). Dark regions represent diminished 
absorption features of the line profiles, interpreted as spotted regions.}
\label{rt_fig}
\end{figure}

To explore the secondary's signatures of stellar activity,
we used the method of Roche tomography 
\citep[RT; ][]{rutten+dhillon94-1,watson+dhillon01-1}, applying an image 
reconstruction technique to obtain a brightness distribution of the secondary 
star. This method is based on the maximum entropy regularisation technique
\citep{skilling+bryan84-1}, which searches for the most featureless
brightness distribution consistent with the data.
Modern RT uses least square deconvolution (LSD) to combine a series of stellar
absorption lines to produce one profile of higher S/N 
\citep[e.g. ][ and references therein]{kochukhovetal10-1,watsonetal06-1}. 
A list of lines was generated using the Vienna Atomic Line 
Database \citep{piskunovetal95-1,kupkaetal99-1}. We used stellar parameters
for an M5V secondary, i.e.\, T$_{\rm eff} = 3~000$ K and $\log g = 4.9$
\citep{baraffeetal98-1}. Finally, to extract the combined line profile
we also need to subtract the continuum from the spectra. For that, we employed 
an iterative procedure masking out the line regions, based on the line list, 
and an estimated line width. For the LSD extraction of the line profile, we 
used the wavelength range 8000-9000 \AA, resulting in a total of 
50 available lines (with line depths greater than 0.5). The resulting line 
profiles have a velocity resolution of $\sim 4$ km/s and signal-to-noise ratio
of about 40-50. A radial-velocity-corrected trailed spectrum of all 
extracted profiles is shown in the top panel of Fig. \ref{lsd_fig}.

The basic input parameters for the RT are: the orbital period, the system's 
inclination, and the masses $M_\mathrm{WD}$ and $M_\mathrm{MS}$ of the stellar
components. In addition, for the non-Roche-lobe-filling secondary star in
LTT 560, and PCEBs in general, its radius $R_\mathrm{MS}$ is needed.
In \citetalias{tappertetal07-1}, we found a large mismatch between the radius
corresponding to the secondary's surface brightness calculated from the 
calibrated spectroscopic data and the radius implied by the photometric light 
curve. We therefore derive that parameter independently of the present data 
by performing a number of RT simulations with all parameters besides
$R_\mathrm{MS}$ fixed, and determining the minimum of the resulting $\chi^2$. 
With this method, we obtain a radius of $R_\mathrm{MS} = 0.16 ^{+0.06}_{-0.03}~
R_\odot$ ($ = 1.1^{+0.4}_{-0.2} \times 10^{10}$ cm), which lies in-between
the two possible radii of $0.8$ and $1.6 \times 10^{10}$ cm from
\citetalias{tappertetal07-1}.

We furthermore included gravity and limb-darkening corrections in the imaging 
procedure, using the coefficients from \citet{claret00-1}. Finally, the
instrumental broadening of the absorption lines was accounted for by
convolving the line profile with a Gaussian with full width at half maximum 
corresponding to the spectral resolution. Following \citet{watsonetal07-1}, we 
do not adopt a two-temperature model \citep{collier-cameron+unruh94-1}. As 
with the Doppler
tomography, only line profiles before the occurrence of the flare were used. 
The resulting surface brightness distribution of 
LTT 560 is shown in Fig. \ref{rt_fig} at five different orientations. 
Brighter regions represent the ``undisturbed" 
photosphere of the star and darker regions are spot-filled areas.

The most prominent feature of the surface brightness distribution is a large 
asymmetric spotted region covering most of the surface of the secondary star.
This may be caused by one large spot covering a considerable 
fraction of the star, as e.g.\,observed by \citet{strassmeier99-1} in the
K0 giant HD 12545, or from a large group of smaller spots, with our data
not allowing to discern between the two scenarios. The observed asymmetry in 
the brightness profile is such that the back face of the secondary is less 
covered with star-spots (brighter) than the rest of the star, with a 
slight tilt in the direction of the stellar rotation. This is consistent with 
our result of Section \ref{flare_sec}, i.e.\,that the leading side of the 
secondary appears to be the more active one.

\section{Discussion}

\begin{figure}
\centering
\includegraphics[angle=270,width=1.0\columnwidth]{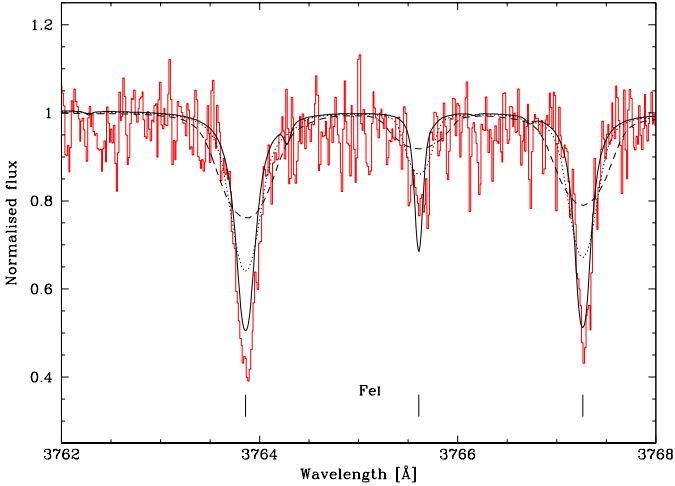}
\caption[]{\label{f-vsini} A close-up of three Fe\,I absorption lines in the 
white dwarf photosphere. Overplotted on the average spectrum are three 
TLUSTY/SYNSPEC models for $T_\mathrm{WD}=7500$\,K, $\log g=7.75$, and the 
abundances listed in Table\,\ref{t-abundances}. The solid line shows a 
non-rotating white dwarf, the dotted line corresponds to $v\sin i=15$\,km/s, 
and the dashed line to $v\sin i=30$\,km/s. }
\end{figure}

We have used Echelle spectroscopy of the PCEB LTT 560 to derive precise
system parameters. From the 
metal abundances in the white dwarf photosphere, we determined an accretion
rate of $\sim5\times10^{-15}\Msun\mathrm{yr^{-1}}$.
\citet{debes06-1} carried out a similar study of six white dwarf and main 
sequence binaries, both short-period (post-common envelope binaries, PCEBs) and
wide binaries, and found accretion rates in the range
$2\times10^{-18}$ to $6\times10^{-16}\Msun\mathrm{yr^{-1}}$. Their study
includes the system RR\,Cae, a white dwarf plus M-dwarf binary that
resembles LTT\,560 in many aspects. Both systems have nearly identical
white dwarf masses and similar companion star masses, but the
orbital period of RR\,Cae is nearly double that of LTT\,560.
\citet{debes06-1} determined an accretion rate of 
$4\times10^{-16}\Msun\mathrm{yr^{-1}}$ for RR\,Cae, and inferred, assuming a
spherically symmetric wind and Bondi-Hoyle type accretion, a wind mass-loss 
rate of $\dot M_\mathrm{dM}=6\times10^{-15}\,\mathrm{M_\odot\,yr^{-1}}$ for
the companion star. Taking the Bondi-Hoyle scenario at face
value, $\dot M_\mathrm{acc}\propto1/R^2$, where $R$ is taken as the
distance between the white dwarf and the M dwarf. For an equally
large wind mass-loss rate, the accretion rate in LTT\,560 would be
about three times higher than in RR\,Cae, which is only a factor of about 
three below the accretion rate that we estimated from the photospheric
abundances. Considering all the uncertainties involved in estimating
the accretion rate and mass-loss rate, RR\,Cae and LTT\,560 appear to
behave rather similar, and we speculate that it is the higher
accretion rate in LTT\,560 that is responsible for the
chromosphere/corona around the white dwarf that we observe in
H$\alpha$. 

The accretion luminosity implied by 
$\dot M_\mathrm{acc}\simeq5\times10^{-15}\Msun\mathrm{yr^{-1}}$ is
$L\simeq1.8\times10^{28}\mathrm{erg\,s^{-1}}$. Integrating the H$\alpha$ flux, 
we find that
$F(\mathrm{H}\alpha)\simeq1.0\times10^{-15}\mathrm{erg\,cm^{-2}\,s^{-1}}$,
or, adopting $d=33$\,pc \citepalias{tappertetal07-1}, 
$L(\mathrm{H}\alpha)\simeq1.3\times10^{26}\mathrm{erg\,s^{-1}}$,
i.e.~it is clear that the accretion-heated layer of the white dwarf
must cool through additional emission mechanisms. In 
\citetalias{tappertetal07-1}, we
discussed how identifying LTT\,560 with a nearby faint
ROSAT\,PSPC source would imply an X-ray luminosity of
$\sim6\times10^{27}\mathrm{erg\,s^{-1}}$, which is comparable to the
predicted accretion luminosity. A deep X-ray observation would be
desirable to confirm the association of LTT\,560 with the ROSAT
X-ray source and to establish more accurately the flux and spectral
shape of the X-ray emission.

The nearly identical $\gamma$ velocities of both the white dwarf's H$\alpha$ 
emission and the photospheric metal absorption lines suggest that they
originate in the same region. However, because of the uncertainties involved 
we cannot exclude that the H$\alpha$ emission is produced somewhat
above the white dwarf's photosphere. For a ``worst case scenario", we
calculate the largest possible difference within the 3$\sigma$ uncertainties
as
\begin{displaymath}
\Delta \gamma = \gamma_\mathrm{WD,metal} + 3\sigma_\mathrm{metal} 
- (\gamma_\mathrm{WD,H\alpha}-3\sigma_\mathrm{H\alpha}) 
= 2.15~\mathrm{km~s^{-1}}.
\end{displaymath}
Since the white dwarf radius is given by
$R_\mathrm{WD} = 0.636~M_\mathrm{WD}/v_\mathrm{gr}$ for $M$ and $R$ in solar 
units and $v_\mathrm{gr}$ in $\mathrm{km~s^{-1}}$, $\Delta \gamma$ translates
into $0.11 R_\mathrm{WD}$. The H$\alpha$ emission could thus in principle 
originate in a region up to $\sim$1000 km above the photosphere. This still 
practically excludes a ``strong" shock as the mechanism behind the emission, 
because the associated shock height is inversely correlated with the 
mass-transfer rate \citep[e.g., Eq.14 in][]{fischer+beuermann01-1}, 
which for LTT\,560 is two orders of magnitude below that of even the 
low-accretion-rate polars \citep{schwopeetal02-1}. It appears more likely 
that the emission line is produced by heat deposited by the accretion 
resulting in a temperature reversal shortly above the white dwarf 
photosphere. However, a detailed treatment of the physics involved is beyond 
the scope of this paper.

The narrow metal lines allow us to place a limit on the
rotation rate of the white dwarf. Fig.\,\ref{f-vsini} shows a close-up
of three Fe\,I lines, along with models for $v\sin i=0$, 15, and
30\,km/s, and suggests that the white dwarf in LTT\,560 is a very slow
($\la15$\,km/s) rotator, similar to the majority of single white
dwarfs \citep{koesteretal98-2, karletal05-1, bergeretal05-2}. This is
an interesting result, as it suggests that the evolution of angular
momentum of white dwarfs in PCEBs is similar to that of field white
dwarfs, with current theories requiring magnetic torques to explain
the observed low rotation rates \citep[e.g.][]{suijsetal08-1}.

The data sets used in \citetalias{tappertetal07-1} and the
present one show the occurrence of flares, indicating that there is an
active secondary star in the system. Roche tomography shows an asymmetric
surface brightness distribution, which we interpret as the presence of star
spots (Fig. \ref{rt_fig}). These high latitude and/or polar spots are very 
common in single active stars \citep[e.g., ][]{strassmeieretal03-1}, and have 
also been observed on donor stars of CVs \citep{watsonetal06-1,watsonetal07-1},
as well as in the pre-CV V471 Tau \citep{hussainetal06-1}. The
models of \citet{granzeretal00-1} suggest that star-spots emerge preferentially
from the flux tube close to the equator of the star and are dragged by the
Coriolis force to the pole of the star. Thus, high latitude and polar spots 
should be a common feature on rapid rotators, such as close binaries. 
However, we caution that these models are only valid for stars 
in the range $0.4~M_\odot \leq M \leq 1.7~M_\odot$ that contain a radiative 
core and a convective shell, and in LTT 560 the M5--6V secondary with a mass 
$M_\mathrm{MS} = 0.14~M_\odot$ is well below this range and can be expected
to be fully convective.

In addition to the high latitude features, we also observe that the inner face 
of the star in our model (the one facing the primary) is slightly darker 
than its backside. The low latitude star-spots populating the region around 
L1 appear to be another common feature in the tidally distorted late-type 
secondaries of 
close binaries \citep{watsonetal06-1,watsonetal07-1,hussainetal06-1}. It has 
been suggested by \citet{holzwarth+schuessler03-1} that tidal interaction 
may force spots to appear at preferred locations. \citet{king+cannizzo98-1} 
find the appearance of such spots at L1 to be the probable mechanism
behind the low brightness states and mass-transfer variations in CVs. As
\citet{watsonetal07-1} point out, a similar effect to that of star spots
at L1 can be achieved by irradiation from the white dwarf primary, although
the low temperature of the white dwarf
\citepalias[$T_\mathrm{WD} = 7500~\mathrm{K}$; ][]{tappertetal07-1} indicates
that this effect will be small if important at all.

What kind of future awaits LTT 560? Following 
\citet{schreiber+gaensicke03-1}\footnote{As already noted by 
\citet{zorotovicetal10-3}, equation (11) in \citet{schreiber+gaensicke03-1} is
incorrect in that the factor $9\pi$ has to be replaced by $2\pi$. Since their
results have been found to be correct, this is merely a typographical error.},
we calculate the period $P_\mathrm{sd}$ at which the system becomes 
semi-detached as
\begin{equation}
P_\mathrm{sd} = 2\pi \left(\frac{R_\mathrm{MS}^3}{G~M_\mathrm{MS}~(1+q^{-1})
                      (R_L/a)^3}\right)^{\frac{1}{2}},
\end{equation}
taken from \citet{ritter86-1}, with the approximation of \citet{eggleton83-1}
\begin{equation}
R_L/a = \frac{0.49~q^\frac{2}{3}}{0.6~q^\frac{2}{3} + \ln (1+q)^\frac{1}{3}}.
\end{equation}
With the parameters given in Table \ref{syspars_tab} (taking the average of 
the He- and the CO-configuration) and the secondary's
radius estimated from the RT, we find that 
$P_\mathrm{sd} = 1.52^{+0.85}_{-0.43}~\mathrm{h}$. 
The large uncertainty here is dominated by our insufficient knowledge of 
$R_\mathrm{MS}$. Modelling the light curve of a photometric time-series data 
set of LTT 560 of high S/N would be desirable to improve the accuracy. 
Determining the distance to the system via a parallax measurement would 
similarly yield the absolute luminosities of the stellar components, thus
provide independent access to both $R_\mathrm{MS}$ and $R_\mathrm{WD}$
and the associated parameters. Nevertheless, a $P_\mathrm{sd}$ below the 
period gap is consistent with the M5--6V spectral type of the secondary star
\citep{beuermannetal98-1,smith+dhillon98-1}. Since it can be assumed that
the secondary is fully convective, we use Eq.8 from 
\citet{schreiber+gaensicke03-1} for angular momentum loss dominated by
gravitational radiation to calculate the time it will take LTT 560 to start 
mass transfer via Roche-lobe overflow to $\sim$3.5 Gyrs. This is much less 
than the Hubble time, thus LTT 560 can be regarded as representative of the 
progenitors of todays CVs. Since the system contains a non-magnetic 
white dwarf and the mass ratio $q < 0.33$, it is likely that the future 
CV LTT 560 will belong to the SU UMa subclass of dwarf novae.

\begin{acknowledgements}
We thank the anonymous referee for helpful comments. Many thanks also to
Matthias Schreiber and Alberto Rebassa for enlightening discussions.

This work has made intensive use of the SIMBAD database, operated at CDS, 
Strasbourg, France, and of NASA's Astrophysics Data System Bibliographic 
Services. IRAF is distributed by the National Optical Astronomy Observatories.
\end{acknowledgements}


\end{document}